\documentclass[12pt]{article}

\usepackage[utf8]{inputenc}
\usepackage[T1]{fontenc}
\usepackage{amsmath,amssymb,amsfonts}
\usepackage{mathtools}
\usepackage{bm}
\usepackage{graphicx}
\usepackage{epstopdf}
\usepackage[dvipsnames]{xcolor}
\usepackage{booktabs}
\usepackage{multirow}
\usepackage{array}
\usepackage{tabularx}
\usepackage{longtable}
\usepackage{geometry}
\usepackage{setspace}
\usepackage{natbib}
\usepackage[colorlinks=true,citecolor=blue,linkcolor=blue,urlcolor=blue]{hyperref}
\usepackage{float}
\usepackage{caption}
\usepackage{enumitem}
\usepackage{threeparttable}

\newcommand{\Ccopula}{\mathcal{C}}
\newcommand{\ccopula}{c}
\newcommand{\Pspace}{\mathcal{P}}

\title{\textbf{Bootstrap-Aggregated Method-of-Moments Estimation of the Copula Correlation Parameter for Marginal Survival Inference under Dependent Censoring}}

\author{
Hyun-Soo Zhang$^{1,2}$, Inkyung Jung$^{1,2}$, Chung Mo Nam$^{*\,1,3,4}$ \\[12pt]
{\small $^{1}$ Division of Biostatistics, Department of Biomedical Systems Informatics,}\\
{\small College of Medicine, Yonsei University, Seoul, Korea}\\
{\small $^{2}$ Biostatistics Collaboration Unit, Department of Biomedical Systems Informatics,}\\
{\small College of Medicine, Yonsei University, Seoul, Korea}\\
{\small $^{3}$ Department of Preventive Medicine, College of Medicine, Yonsei University, Seoul, Korea}\\
{\small $^{4}$ Graduate School of Public Health, Yonsei University, Seoul, Korea}
}

\date{}

\begin{document}

\maketitle

\begin{abstract}
In dependently censored survival data, the usual assumption of independent censoring or an incorrect specification of the correlation between the event and censoring times can bias marginal survival inference. Likelihood-based estimation of this dependence can be numerically unstable with large variance, and practical alternatives are limited. The proposed method uses generalized method-of-moments to estimate the copula correlation parameter of a Normal, Clayton, Gumbel, or Frank copula that links exponential, Weibull, or log-normal marginal survival times. Bootstrap-aggregation of simulated annealing is employed over candidate correlation ranges to obtain stable estimates. Simulations assess accuracy and uncertainty via mean absolute error, bootstrap confidence intervals, and empirical coverage. The method is applied to a double-blind randomized clinical trial with dependent censoring from early patient dropouts, where accurate marginal survival inference is needed to estimate the effect of a treatment on patient survival.
\end{abstract}


\section{Introduction}

Right-censoring (``Censoring'' hereafter) is a central topic in survival analysis, where independent or non-informative censoring is an important assumption for estimating and comparing marginal survival functions. In the biomedical context of a randomized clinical trial (RCT), the primary endpoint may be the patient's overall survival (OS) time under a new drug versus a control drug. Commonly observed in oncology trials, patients may experience toxicity or deteriorating health and withdraw from the trial, resulting in censoring times. Assuming independent censoring to compare the survival probabilities of the two drug groups is doubtful here, as censoring due to worsening health would be positively correlated to the patient's OS. Applying the Kaplan-Meier (K-M) method that assumes independent censoring for the marginal survival of the new drug group would over-estimate the true survival probability, and statistical approaches to deal with dependent censoring are thus needed~\cite{Emura2018,Klein2010,Moeschberger1995}.

Dependent censoring also includes competing risks survival data, where the occurrence of an event prevents other events from occurring~\cite{Putter2007,Crowder1994,Prentice1978}. Such situations frequently arise in the biomedical setting, where death from other causes is a competing risk to death from cancer, or death without relapse is a competing risk to disease relapse in leukemia patients. When the competing event may be associated with the event of interest, it represents dependent censoring. In this ``classical'' competing risks situation (in contrast to ``semi-'' competing risks, such as the death without relapse and disease relapse example above), only one event and its event time are observable. Here, we focus on classical competing risks where the occurrence of a competing event results in right-censoring of the event of interest and vice versa. We also mainly consider bivariate survival times of 1) the time-to-event of interest (``event'' time), and 2) the time to all other competing or censoring events (``censoring'' time). Among the different types of hazard functions in competing risks survival analysis~\cite{Emura2018,Andersen2012}, we focus on the ``marginal'' hazard function, which includes the well-known ``cause-specific'' hazard as a special case. In fact, the cause-specific hazard is a marginal hazard assuming independence as its correlation structure. It represents one of many possible marginal hazards determined by the correlation between the two survival times. The ``sub-distributional'' hazard, made popular by the Fine \& Gray model~\cite{Fine1999}, is another type of hazard function widely employed in the competing risks literature but is not further pursued here. Distinguishment between the three types of hazard functions, as well as their interrelationships, are well described in the paper by Emura et al.~\cite{Emura2020}.

Under dependent censoring, accurate estimation of the correlation between event and censoring times is critical for valid inference on marginal survival functions and treatment effects. A fundamental problem is that only the minimum of the two survival times is observed and never both, which is known as the non-identifiability of competing risks~\cite{Tsiatis1975,Cox1959}. Additional information of the correlation structure is thus needed and various efforts such as using informative covariates or assuming parametric distributions continue today~\cite{Schneider2023,Jo2023,Wang2023,Deresa2024,Emura2016,Czado2023,Sorrell2023,Gharari2023}. A subtle yet important distinction to make here is whether identifiability is being defined nonparametrically or parametrically~\cite{Jo2023,Rivest2001}. The widely cited non-identifiability proof by Tsiatis~\cite{Tsiatis1975} did not make any parametric assumptions regarding the marginal survival times, and the non-identifiability proven was thus ``nonparametric''. In contrast, more recent efforts~\cite{Schneider2023,Jo2023,Wang2023,Deresa2024,Gharari2023}, including this study, utilize a ``parametric'' approach in identifying the correlation for dependently censored survival data. Although the marginal survival times cannot be identified nonparametrically, selective parametric restrictions enable such identification.

A useful tool for dependence modeling is the copula function, where a copula is a multivariate CDF with uniformly distributed marginals~\cite{Nelsen2006,Sklar1959}. From Sklar's theorem~\cite{Sklar1959}, a copula determines the dependence among the components of a random vector, for which the book by Nelsen~\cite{Nelsen2006} provides comprehensive details. A major application of copulas in survival analysis is the construction of correlated survival times, and they have been widely adapted in dependence modeling of competing risks survival data. Zheng and Klein~\cite{Zheng1995,Zheng1996} have done seminal work in this area, where they proposed an ``assumed'' copula to first determine the dependence structure, followed by unbiased identification of the marginal hazard functions. Huang and Zhang~\cite{Huang2008} applied Zheng and Klein's work to a Cox proportional hazards (PH) regression setting for sensitivity analyses based on a plausible range of correlations between the event and censoring times.

More systematically, Chen~\cite{Chen2010} extended the assumed copula approach by generalizing the marginal regressions via semiparametric transformation models. Here, both the functional form of the copula and its correlation parameter were assumed as known. Chen commented that the focus on ``marginal'' or ``net'' event time regression analysis implies the hypothetical setting of artificially removing the other competing risk, i.e., the assessment of covariate effects on either one of the bivariate survival times but having the other correlated event time removed. This is also a main motivation of the current study, such as estimating the marginal effect of a new drug on OS if no patients were dependently censored owing to early withdrawal from an RCT. Chen also noted that in the presence of regression covariates, although the copula correlation parameter may be estimated from the data along with the regression coefficients~\cite{Braekers2005}, the variability of the resulting estimates is large and the estimates are unstable. Emura and Chen~\cite{Emura2016} applied Chen's copula-based framework to gene selection for survival data with dependent censoring, where they noted that no practical methods of simultaneously estimating the copula association parameter and the marginal regression models yet exist. An important comment was that due to the non-identifiability of competing risks data, the usual likelihood equation may provide little information about the true association parameter or Kendall's tau correlation, thus rescinding maximum likelihood-based approaches. An implementation of their work is provided as an R package \textit{compound.Cox}~\cite{Emura2019}, which was also used in our analyses.

Using parametrically specified copulas and marginal distributions, Van Keilegom et al.~\cite{Deresa2024,Czado2023,Deresa2022ins,Deresa2020csda,Deresa2021,Deresa2020bj} have extensively studied the identifiability of dependent competing risks data. Their recent work~\cite{Czado2023} considerably expands the parametric approach to marginal distributions widely used in practice. An event time $T$ that is stochastically dependent on a censoring time $C$ is considered, where the interest is in estimating the marginal distribution of $T$. They provided sufficient conditions where a parametric copula and parametric marginal distributions of $T$ and $C$ are identifiable without assuming that the correlation parameter is known. Although the study covers often used survival time distributions and copulas, some distributions such as the Gompertz do not satisfy this identifiability condition.

Other recent works on dependent censoring are by Wang, Jo et al., Schneider et al., and Gharari et al.~\cite{Schneider2023,Jo2023,Wang2023,Gharari2023}. Wang showed that the correlation structure becomes identifiable in bivariate dependent competing risks data $(T, C)$ if the distribution of $T$ or $C$ is exponential, and Jo et al.\ used the method-of-moments to estimate the copula correlation parameter, assuming that an informative covariate related to both the event and censoring times exists. Schneider et al.\ used a Clayton copula to model dependently censored survival data. They employed a Bayesian approach, stating that non-informative priors can estimate the copula correlation parameter whereas maximum likelihood inference requires the correlation parameter to be predefined or known. Finally, Gharari et al.\ utilized a deep learning-based approach while assuming the copula function to be known under a Weibull Cox model.

Following these developments in copula dependence modeling, the current study proposes a new method to identify the correlation parameter in dependently censored survival data. Specifically, a broad parametric framework encompassing exponential, Weibull, and log-normal distributions is utilized to estimate the correlation between bivariate survival times. To overcome the aforementioned limitation of maximum likelihood estimation (MLE) in simultaneously estimating the copula correlation and marginal distribution parameters~\cite{Schneider2023,Emura2016,Chen2010,Michimae2022}, method-of-moments bootstrap aggregating (bagging), in conjunction with the copula-graphic estimator of marginal survival, is proposed. After a simulation study on the performance of the proposed method, we demonstrate its application in treatment effect estimation with a simulated RCT. The proposed method is also applied to a real-world, double-blind RCT among adults infected with the human immunodeficiency virus (HIV)~\cite{Hammer1996,Juraska2022}, focusing on the correlation between the time to the primary endpoint and the time to early withdrawal.

The remainder of this paper is organized as follows: Section~2 introduces the proposed method of correlation estimation in dependently censored survival data. Section~3 presents a simulation study evaluating the performance of the proposed method across various correlations and marginal distributions, including a comparison with maximum likelihood estimation (MLE) results. Section~4 applies the proposed method to a real-world HIV dataset and compares its results with those from previous studies. Section~5 discusses and concludes the study.

\section{Proposed Method}

\subsection{Set-up and notation}

Let $T$: event time and $C$: dependent censoring time. For subject $i = 1, 2, \ldots, n$, we observe
\begin{equation}
X_{i} = \min\!\left(T_{i}, C_{i}\right),\quad \delta_{i} = \mathbb{I}\!\left\{T_{i} \leq C_{i}\right\}.
\end{equation}
Here, $X_{i}$ is the observed follow-up time and $\delta_{i}$ is the event indicator. Let $F_{T}$, $F_{C}$ be the marginal distribution functions (CDFs), $f_{T}$, $f_{C}$ their densities (PDFs), and $S_{T}$, $S_{C}$ their survival functions.

Assume a parametric copula $\Ccopula\colon [0,1]^{2} \to [0,1]$,
\begin{equation}
\Ccopula(u, v) = \Pr(U \leq u,\, V \leq v),\quad U \sim \text{Uniform}(0,1),\; V \sim \text{Uniform}(0,1),
\end{equation}
that captures the dependence structure and links the marginals into a joint distribution~\cite{Nelsen2006,Sklar1959} as
\begin{equation}
H(t, c) = \Pr(T \leq t,\, C \leq c) = \Ccopula\!\left(F_{T}(t),\, F_{C}(c)\right).
\end{equation}

The copula derivatives are written as
\begin{equation}
\frac{\partial}{\partial u}\Ccopula(u,v) = \Pr(V \leq v \mid U = u) = \Ccopula_{u}(u,v),\quad
\frac{\partial^{2}}{\partial u\,\partial v}\Ccopula(u,v) = \ccopula(u,v),
\end{equation}
where $\ccopula(u,v)$ may not always exist~\cite{Nelsen2006}.

This study considers two major families of copulas: Archimedean and Elliptical. Archimedean copulas, such as the Clayton, Gumbel, and Frank, are defined by a generator function $\phi$. Elliptical copulas, such as the Gaussian or normal copula, are derived from multivariate elliptical distributions. The strength of the correlation between $T$ and $C$ is quantified using Kendall's tau, a nonparametric rank-based correlation~\cite{Nelsen2006,Sklar1959} (Supplementary Table~S1).

Parametric marginals of exponential, Weibull, or log-normal distributions are considered for $T$ and $C$~\cite{Collett2023,Klein2003}. Closed-form expressions for the three marginals as special cases of the generalized gamma density are provided as Supplementary material.

In the case of log-normal marginals, the full parameter vector is
\begin{equation}
\theta = \left(\mu_{T},\, \mu_{C},\, \sigma_{T}^{2},\, \sigma_{C}^{2},\, \rho\right),
\end{equation}
where $(\mu_{T}, \sigma_{T}^{2})$ and $(\mu_{C}, \sigma_{C}^{2})$ parameterize the two log-normal marginals, and $\rho$ denotes Kendall's tau between $\log(T)$ and $\log(C)$, which is invariant under monotonic transformations. Since the exponential distribution is a special case of the Weibull, specifying two parameters (shape, scale) for each marginal distribution of $T$ and $C$ plus one correlation parameter also yields a 5-dimensional parameter vector.

\subsection{Distribution of \texorpdfstring{$(X,\,\delta)$}{(X, delta)}}

While $\Ccopula(u,v)$ is a joint CDF, we are more interested in the joint survival function expressed as $\Ccopula\!\left(S_{T}(t), S_{C}(c)\right) = \Pr(T > t, C > c)$. The survival function of $X$ is thus
\begin{equation}
S_{X}(x) = \Pr(X > x) = \Pr(T > x, C > x) = \Ccopula\!\left(S_{T}(x), S_{C}(x)\right).
\end{equation}

The joint probabilities of $(X, \delta)$ for continuous $T$ and $C$ can be written as
\begin{align}
\Pr(\delta = 1,\, X \in dx) &= f_{T}(x) \cdot \Pr(C \geq x \mid T = x)\,dx, \\
\Pr(\delta = 0,\, X \in dx) &= f_{C}(x) \cdot \Pr(T \geq x \mid C = x)\,dx.
\end{align}

Utilizing copula notation,
\begin{align}
\Pr(C \geq x \mid T = x) &= 1 - \Ccopula_{u}\!\left(F_{T}(x), F_{C}(x)\right), \\
\Pr(T \geq x \mid C = x) &= 1 - \Ccopula_{v}\!\left(F_{T}(x), F_{C}(x)\right),
\end{align}
where $\Ccopula_{u} = \frac{\partial \Ccopula}{\partial u}$, $\Ccopula_{v} = \frac{\partial \Ccopula}{\partial v}$.

The probability of observing the event of interest (survival time $T$) is then
\begin{equation}
p(\theta) = \Pr(\delta = 1) = \int_{0}^{\infty} f_{T}(x) \cdot \left\{1 - \Ccopula_{u}\!\left(F_{T}(x), F_{C}(x)\right)\right\}\,dx.
\end{equation}

\subsection{Moment functions and generalized method-of-moments (GMM) estimation}

For the problem of dependent censoring, it has been established that the bivariate normal and bivariate Weibull distributions are identifiable from the observable $(X, \delta)$~\cite{Nadas1971,Basu1978,Moeschberger1974,Emoto1990}. Thus, equating a set of population moments to their corresponding sample moments yields an estimate for the parameter vector $\theta$. GMM estimation, formalized by Hansen~\cite{Hansen1982} and comprehensively treated by Hall~\cite{Hall2003}, does not require complete knowledge of the underlying distribution and can be computationally less burdensome than MLE. The moment functions and GMM estimation in the case of log-normal marginals are as follows.

Working on the log scale,
\begin{equation}
Y_{1} = \log(T),\quad Y_{2} = \log(C),\quad W = \log(X) = \min(Y_{1}, Y_{2}),\quad \text{and}\quad \delta = \mathbb{I}\{Y_{1} \leq Y_{2}\}.
\end{equation}

The sample moments are defined as
\begin{equation}
\widehat{p} = \frac{1}{n}\sum_{i=1}^{n} \delta_{i},
\end{equation}
\begin{equation}
\widehat{\mu}_{1} = \frac{\sum_{i} \delta_{i} \log X_{i}}{\sum_{i} \delta_{i}},\qquad
\widehat{\sigma}_{1}^{2} = \frac{\sum_{i} \delta_{i}\left(\log X_{i} - \widehat{\mu}_{1}\right)^{2}}{\sum_{i} \delta_{i}},
\end{equation}
\begin{equation}
\widehat{\mu}_{2} = \frac{\sum_{i}(1-\delta_{i})\log X_{i}}{\sum_{i}(1-\delta_{i})},\qquad
\widehat{\sigma}_{2}^{2} = \frac{\sum_{i}(1-\delta_{i})\left(\log X_{i} - \widehat{\mu}_{2}\right)^{2}}{\sum_{i}(1-\delta_{i})}.
\end{equation}

Their theoretical counterparts are
\begin{equation}
p(\theta) = E[\delta],
\end{equation}
\begin{equation}
\mu_{1}(\theta) = E\!\left[Y_{1}\;\middle|\; Y_{1} \leq Y_{2}\right],\qquad
\sigma_{1}^{2}(\theta) = \text{Var}\!\left[Y_{1}\;\middle|\; Y_{1} \leq Y_{2}\right],
\end{equation}
\begin{equation}
\mu_{2}(\theta) = E\!\left[Y_{2}\;\middle|\; Y_{2} < Y_{1}\right],\qquad
\sigma_{2}^{2}(\theta) = \text{Var}\!\left[Y_{2}\;\middle|\; Y_{2} < Y_{1}\right].
\end{equation}

Expressions for the moments of a bivariate normal $Y_{1} \sim N(\mu_{T}, \sigma_{T}^{2})$, $Y_{2} \sim N(\mu_{C}, \sigma_{C}^{2})$ with correlation $\rho$ are given using
\begin{equation}
D = Y_{1} - Y_{2} \sim N(m, s^{2}),\qquad m = \mu_{T} - \mu_{C},\qquad s^{2} = \sigma_{T}^{2} + \sigma_{C}^{2} - 2\rho\sigma_{T}\sigma_{C}.
\end{equation}

Additionally, let
\begin{equation}
\kappa = -\frac{m}{s},\qquad
\zeta_{1} = \frac{\varphi(\kappa)}{\Phi(\kappa)},\qquad
\zeta_{2} = \frac{\varphi(\kappa)}{1-\Phi(\kappa)} = \frac{\varphi(\kappa)}{\Phi(-\kappa)},
\end{equation}
where $\varphi$ and $\Phi$ are the standard normal PDF and CDF.

The theoretical moments are then
\begin{equation}
p(\theta) = \Pr(D \leq 0) = \Phi(\kappa),
\end{equation}
\begin{equation}
\mu_{1}(\theta) = \mu_{T} - \frac{\sigma_{T}(\sigma_{T} - \rho\sigma_{C})}{s}\zeta_{1},\qquad
\sigma_{1}^{2}(\theta) = \sigma_{T}^{2} - \left\{\frac{\sigma_{T}(\sigma_{T}-\rho\sigma_{C})}{s}\right\}^{2}\!\left(\kappa\zeta_{1} + \zeta_{1}^{2}\right),
\end{equation}
\begin{equation}
\mu_{2}(\theta) = \mu_{C} - \frac{\sigma_{C}(\sigma_{C} - \rho\sigma_{T})}{s}\zeta_{2},\qquad
\sigma_{2}^{2}(\theta) = \sigma_{C}^{2} - \left\{\frac{\sigma_{C}(\sigma_{C}-\rho\sigma_{T})}{s}\right\}^{2}\!\left(-\kappa\zeta_{2} + \zeta_{2}^{2}\right).
\end{equation}

Derivation of the theoretical moments is provided as Supplementary material.

Defining the sample and population moment vectors as
\begin{equation}
\widehat{\bm{m}} = \left(\widehat{p},\; \widehat{\mu}_{1},\; \widehat{\mu}_{2},\; \widehat{\sigma}_{1}^{2},\; \widehat{\sigma}_{2}^{2}\right)',
\end{equation}
\begin{equation}
\bm{m}(\theta) = \left(p(\theta),\; \mu_{1}(\theta),\; \mu_{2}(\theta),\; \sigma_{1}^{2}(\theta),\; \sigma_{2}^{2}(\theta)\right)',
\end{equation}
the GMM estimator $\widehat{\theta}$ solves $\bm{m}(\theta) = \widehat{\bm{m}}$ by minimizing
\begin{equation}
Q_{n}(\theta) = \left\{\bm{m}(\theta) - \widehat{\bm{m}}\right\}' W_{n} \left\{\bm{m}(\theta) - \widehat{\bm{m}}\right\},
\end{equation}
where $W_{n}$ is a diagonal weight matrix with entries given by the inverse bootstrap variances of $\widehat{\bm{m}}$~\cite{Casella2024,Givens2012}.

\subsection{Feasible parameter space using copula-graphic bounds}

To regularize the search space, we construct a feasible parameter region using the copula-graphic (CG) estimator~\cite{Emura2018,Zheng1995,Zheng1996} of marginal survival functions under several representative values of $\rho$, e.g., correlations of 0 (none), 0.3 (low), 0.5 (moderate), or 0.8 (high) in terms of Kendall's tau. For each representative $\rho$ we compute $\widehat{S}_{T}^{CG}$ and $\widehat{S}_{C}^{CG}$, fit parametric marginals to obtain preliminary estimates $(\widehat{\mu}_{T}, \widehat{\sigma}_{T}^{2}, \widehat{\mu}_{C}, \widehat{\sigma}_{C}^{2})^{CG}$, and take empirical quartile ranges across the preliminary estimates obtained from the representative $\rho$'s to form lower and upper bounds for each of the marginal parameters. This defines a rectangular search region $\Pspace$ for $(\mu_{T}, \sigma_{T}^{2}, \mu_{C}, \sigma_{C}^{2})$, for which additional details are included as Supplementary material.

\subsection{Bootstrap-aggregating (Bagging) based estimation}

To minimize the objective function $Q_{n}(\theta)$, a two-stage estimation procedure is proposed. The first global stage employs bootstrap-aggregating (Bagging)~\cite{Breiman1996} to identify the most plausible range of the correlation parameter, along with the marginal parameter estimates within $\Pspace$. The second stage uses a local search algorithm to refine the parameter estimates within this identified range of $\rho$.

\subsubsection{Global stage}

We use Generalized Simulated Annealing (GenSA)~\cite{Cortez2014,Xiang2017,Xiang2013} for stochastic global optimization, leveraging its ability to escape local minima and converge to the global minimum. GenSA is embedded within a bagging framework that entails voting among four candidate correlation ranges of $(-0.1,\,0.15]$, $(0.15,\,0.4]$, $(0.4,\,0.65]$, and $(0.65,\,0.9]$, corresponding to none, low, moderate, and high correlation, respectively. We assume negative correlations to be dealt with by applying a monotone decreasing transformation (e.g., reciprocal) to one of the time-to-events. The bagged GenSA procedure is as follows:

\begin{enumerate}[label=\arabic*)]
\item \textbf{Bootstrap sampling:} Generate bootstrap replicates $b = 1, \ldots, B$.
\item \textbf{Parallel global search:} For each bootstrap sample, perform four GenSA optimizations simultaneously, where each minimizes $Q_{n}(\theta)$ constrained to one of the four candidate correlation ranges. The marginal parameters are constrained within $\Pspace$.
\item \textbf{Voting:} For a given bootstrap sample, the four optimization runs yield four minimized values of $Q_{n}(\theta)$. The correlation range corresponding to the lowest value of $Q_{n}(\theta)$ receives the vote for that sample.
\item \textbf{Aggregation:} After iterating through $B$ samples, tally the votes for each of the four correlation ranges. The range with the highest vote count is selected as the most plausible correlation range.
\end{enumerate}

\subsubsection{Local stage}

After the correlation range is identified, the parameter vector $\widehat{\theta}$ is obtained through a local search. Standard gradient-based optimization~\cite{Givens2012}, such as the \texttt{nlminb()} function in R, is employed. The search is initialized with the best parameter set found during the global stage and is constrained within the identified correlation range and the marginal parameter region $\Pspace$. The local search iterates until a convergence criterion is met, e.g., decrease in $Q_{n}(\theta) < 10^{-8}$. The final estimate is $\widehat{\theta} = (\widehat{\mu}_{T}, \widehat{\mu}_{C}, \widehat{\sigma}_{T}^{2}, \widehat{\sigma}_{C}^{2}, \widehat{\rho})$, where $\widehat{\rho}$ is the estimated correlation between $T$ and $C$.

\subsection{Large sample properties and inference}

The proposed estimator is a specific application of the GMM framework, for which the large sample properties are well-established~\cite{Hansen1982,Hall2003,Belalia2024,Oh2013}. Under regularity conditions, the estimator $\widehat{\theta}$ minimizing $Q_{n}(\theta)$ is consistent for the true parameter value (denote as $\theta_{0}$) and is asymptotically normally distributed.

The regularity conditions for consistency include:

\begin{enumerate}[label=\arabic*)]
\item Correct specification of the moment conditions, i.e., $E(\widehat{\bm{m}}) = \bm{m}(\theta_{0})$,
\item Identifiability of $\theta_{0}$, i.e., $\bm{m}(\theta) \neq \bm{m}(\theta_{0})$ for $\theta \neq \theta_{0}$,
\item Compactness of the parameter space, i.e., the five-dimensional $\Theta = \Pspace \times [\epsilon_{\rho},\, 1-\epsilon_{\rho}]$, $0 < \epsilon_{\rho} < 1$, is bounded and closed,
\item Continuity of the moment functions.
\end{enumerate}

Regarding the asymptotic distribution of the proposed estimator, if $\bm{m}(\theta)$ is continuously differentiable at $\theta_{0}$ with full-rank derivative $G = \partial \bm{m}(\theta)/\partial\theta'\big|_{\theta_{0}}$, and for $\Sigma$: asymptotic covariance matrix of the sample moments,
\begin{equation}
\sqrt{n}\left(\widehat{\bm{m}} - \bm{m}(\theta_{0})\right) \overset{d}{\to} N(0, \Sigma)
\end{equation}
holds, then the GMM estimator $\widehat{\theta}$ satisfies
\begin{equation}
\sqrt{n}\left(\widehat{\theta} - \theta_{0}\right) \overset{d}{\to} N(0, V),\qquad V = \left(G'WG\right)^{-1}\left(G'W\Sigma WG\right)\left(G'WG\right)^{-1},
\end{equation}
where $W$ is the probability limit of $W_{n}$, and $V$ is the familiar sandwich estimator. The sample analogues $\widehat{\Sigma}$, $\widehat{G}$, and $W_{n}$ may be substituted into the expression for $V$. Our proposed estimator is a continuous function of sample moments, and even if its theoretical moments are not explicitly derived in closed form (i.e., require numerical integration or simulation), asymptotic normality continues to hold provided that the simulation size grows with the sample size~\cite{Belalia2024,Oh2013}.

Estimator uncertainty was quantified using the nonparametric bootstrap in finite-sample simulations. We report Monte Carlo summaries of point-estimation accuracy (bias and mean absolute error; MAE) across simulation replicates and use bootstrap standard errors (SEs) and confidence intervals (CIs) to assess estimator uncertainty and coverage. With Kendall's tau estimates arranged in increasing order $\widehat{\tau}_{(1)} \leq \widehat{\tau}_{(2)} \leq \ldots \leq \widehat{\tau}_{(B)}$ for $B$ bootstrap samples, bootstrap CIs are calculated as $\left[\widehat{\tau}_{\left(B \cdot \frac{\alpha}{2}\right)},\, \widehat{\tau}_{\left(B \cdot (1-\frac{\alpha}{2})\right)}\right]$ by the percentile method~\cite{Givens2012}.

\section{Simulation Study: Estimation of Correlation in Dependently Censored Survival Data}

\subsection{Data generation and simulation parameter settings}

The conditional CDF method via copula partial derivatives~\cite{Sorrell2023,Nelsen2006,Sklar1959} is used to generate correlated bivariate survival times that follow certain marginal distributions. For example, after generating a random variable $U = u \sim \text{Uniform}(0,1)$ and its correlated random variable $V = v \sim \text{Uniform}(0,1)$ through the conditional CDF of a normal copula, consider a bivariate Weibull distribution with marginals $T \sim \text{Weibull}(\alpha_{T}, \lambda_{T})$ and $C \sim \text{Weibull}(\alpha_{C}, \lambda_{C})$ for shape $\alpha$ and scale $\lambda$. From a Weibull distribution's hazard, cumulative hazard, and survival functions
\begin{equation}
h(t) = \alpha\lambda t^{\alpha-1},\quad H(t) = \int_{0}^{t}h(u)\,du = \lambda t^{\alpha},\quad\text{and}\quad S(t) = \exp[-H(t)] = \exp[-\lambda t^{\alpha}],
\end{equation}
the correlated bivariate Weibull survival times of $T$ (= $t$) and $C$ (= $c$) are generated as
\begin{align}
u &= S(t) = \exp[-\lambda_{T} t^{\alpha_{T}}],\quad t = S^{-1}(u) = \left\{-\frac{\log(u)}{\lambda_{T}}\right\}^{1/\alpha_{T}}, \\
v &= S(c) = \exp[-\lambda_{C} c^{\alpha_{C}}],\quad c = S^{-1}(v) = \left\{-\frac{\log(v)}{\lambda_{C}}\right\}^{1/\alpha_{C}}.
\end{align}

The overall simulation configuration is based upon varying the three factors of: a) marginal distributions: exponential, Weibull, or log-normally distributed; b) strength of the correlation: Kendall's tau of 0, 0.3, 0.5, or 0.8; and c) functional form of copulas: normal, Clayton, Frank, or Gumbel copulas.

The results of varying a) and b) under a normal copula are presented first, while the results of varying c) are presented in subsection~3.3 with a comparison to those using MLE.

The marginal distribution parameters follow the settings of previous studies on copula dependence modeling~\cite{Czado2023,Sorrell2023}. Specifically, the exponential distribution scale parameters are set as $T \sim \text{Exponential}(\lambda_{T}=0.025)$, $C \sim \text{Exponential}(\lambda_{C}=0.039)$, from the renal transplant data example of Sorrell et al.~\cite{Sorrell2023}, the Weibull shape and scale parameters are set as $T \sim \text{Weibull}(\alpha_{T}=0.63, \lambda_{T}=0.06)$, $C \sim \text{Weibull}(\alpha_{C}=0.86, \lambda_{C}=0.04)$, also from Sorrell et al., and the log-normal mean and standard deviation parameters are set as $T \sim \text{log-normal}(\mu_{T}=2.2, \sigma_{T}=1.0)$, $C \sim \text{log-normal}(\mu_{C}=2.0, \sigma_{C}=0.25)$, from Czado and Van Keilegom~\cite{Czado2023}.

The normal copula's correlation parameter $\rho$ values are 0, 0.454, 0.7071, and 0.9510, corresponding to Kendall's tau of 0, 0.3, 0.5, and 0.8, respectively (Supplementary Table~S1). The sample sizes of the simulated $(T, C)$ are varied among $n = 100$, 200, 500, 1000, or 2000, where $X_{i} = \min(T_{i}, C_{i})$ and $\delta_{i} = \mathbb{I}\{T_{i} \leq C_{i}\}$ are defined for each row $i = 1, 2, \ldots, n$.

For each simulated $(T, C)$, bootstrap samples of equal size $n$ are used to estimate SEs and 95\% CIs. 200 bootstrap samples are taken to calculate the bootstrap SE and 95\% CI of the point estimate $\widehat{\rho}$. Additionally, multiple runs of $(T, C)$ data generation and subsequent bootstraps are conducted to obtain the MAE and CP of the 95\% CIs. 100 multiple runs of 100 bootstrap samples each are utilized here.

\subsection{Simulation results}

Simulation results of correlation estimation in dependently censored survival data $(T, C)$ using the proposed method are shown in Tables~\ref{tab:table1} and~\ref{tab:table2}.

\begin{table}[H]
\centering
\caption{Simulation results of correlation (dependence) estimation with the proposed method in dependently censored survival data $(T, C)^{\text{a}}$ and its bootstrap samples, where the two marginal survival times are correlated by the normal (Gaussian) copula}
\label{tab:table1}
\begin{threeparttable}
\small
\begin{tabular}{lccccl}
\toprule
Marginal distributions & True $\tau$ & Point Est. & Bootstr.$^{\text{b}}$ MAE & Bootstr. SE & Bootstr. 95\% CI \\
\midrule
 & 0 & 0.016 & 0.075 & 0.053 & $(-0.049,\; 0.147)$ \\
Bivariate & 0.3 & 0.393 & 0.054 & 0.060 & $(0.188,\; 0.398)$ \\
exponential & 0.5 & 0.570 & 0.071 & 0.065 & $(0.418,\; 0.645)$ \\
 & 0.8 & 0.818 & 0.054 & 0.050 & $(0.727,\; 0.898)$ \\
\midrule
 & 0 & 0.111 & 0.060 & 0.070 & $(-0.094,\; 0.142)$ \\
Bivariate & 0.3 & 0.369 & 0.056 & 0.065 & $(0.175,\; 0.391)$ \\
Weibull & 0.5 & 0.444 & 0.064 & 0.073 & $(0.406,\; 0.646)$ \\
 & 0.8 & 0.816 & 0.048 & 0.054 & $(0.701,\; 0.895)$ \\
\midrule
 & 0 & $-0.098$ & 0.059 & 0.070 & $(-0.096,\; 0.137)$ \\
Bivariate & 0.3 & 0.356 & 0.059 & 0.069 & $(0.171,\; 0.394)$ \\
log-normal & 0.5 & 0.570 & 0.067 & 0.068 & $(0.414,\; 0.645)$ \\
 & 0.8 & 0.798 & 0.062 & 0.070 & $(0.659,\; 0.895)$ \\
\midrule
 & 0 & $-0.093$ & 0.055 & 0.053 & $(-0.095,\; 0.109)$ \\
Weibull \& & 0.3 & 0.276 & 0.076 & 0.059 & $(0.154,\; 0.360)$ \\
log-normal & 0.5 & 0.524 & 0.054 & 0.049 & $(0.402,\; 0.558)$ \\
 & 0.8 & 0.732 & 0.063 & 0.057 & $(0.660,\; 0.845)$ \\
\bottomrule
\end{tabular}
\begin{tablenotes}[flushleft]\footnotesize
\item Abbreviations: Bootstr., bootstrap; CI, confidence interval; Est., estimate; MAE, mean absolute error; SE, standard error.
\item[$^{\text{a}}$] Sample sizes were 100, 200 or 500 for Kendall's tau of 0 or 0.8, and 1000 or 2000 for Kendall's tau of 0.3 or 0.5.
\item[$^{\text{b}}$] 200 bootstrap samples were used.
\end{tablenotes}
\end{threeparttable}
\end{table}

\begin{table}[H]
\centering
\caption{Simulation results of correlation (dependence) estimation with the proposed method in multiple generations of dependently censored survival data $(T, C)^{\text{a}}$ and its bootstrap samples, where the two marginal survival times are correlated by the normal (Gaussian) copula}
\label{tab:table2}
\begin{threeparttable}
\small
\begin{tabular}{lccccc}
\toprule
Marginal distributions & True $\tau$ & Mean Est. & MAE$^{\text{b}}$ & Empirical SE & CP \\
\midrule
 & 0 & 0.028 & 0.068 & 0.088 & 96 \\
Bivariate & 0.3 & 0.295 & 0.061 & 0.077 & 97 \\
exponential & 0.5 & 0.523 & 0.061 & 0.070 & 96 \\
 & 0.8 & 0.805 & 0.062 & 0.075 & 95 \\
\midrule
 & 0 & 0.061 & 0.077 & 0.081 & 96 \\
Bivariate & 0.3 & 0.262 & 0.069 & 0.075 & 92 \\
Weibull & 0.5 & 0.558 & 0.074 & 0.068 & 96 \\
 & 0.8 & 0.815 & 0.066 & 0.077 & 97 \\
\midrule
 & 0 & 0.013 & 0.063 & 0.083 & 98 \\
Bivariate & 0.3 & 0.287 & 0.055 & 0.060 & 94 \\
log-normal & 0.5 & 0.492 & 0.054 & 0.049 & 96 \\
 & 0.8 & 0.755 & 0.075 & 0.081 & 96 \\
\midrule
 & 0 & 0.021 & 0.068 & 0.082 & 93 \\
Weibull \& & 0.3 & 0.286 & 0.070 & 0.090 & 92 \\
log-normal & 0.5 & 0.493 & 0.053 & 0.068 & 94 \\
 & 0.8 & 0.752 & 0.067 & 0.072 & 97 \\
\bottomrule
\end{tabular}
\begin{tablenotes}[flushleft]\footnotesize
\item Abbreviations: CP, coverage probability; Est., estimate; MAE, mean absolute error; SE, standard error.
\item[$^{\text{a}}$] Sample sizes were 100, 200 or 500 for Kendall's tau of 0 or 0.8, and 1000 or 2000 for Kendall's tau of 0.3 or 0.5.
\item[$^{\text{b}}$] 100 multiple runs (generations) of 100 bootstrap samples were used.
\end{tablenotes}
\end{threeparttable}
\end{table}

Table~\ref{tab:table1} shows the accuracy of the proposed method in terms of the point estimate and bootstrap SE of the true correlation in a simulated $(T, C)$ dataset. The proposed method showed accurate point estimates of $\rho$ under the widely used survival time distributions of exponential, Weibull, or log-normal. The bootstrap 95\% CIs also demonstrate the proposed method's capability of distinguishing between none, low, moderate, or high correlation between $T$ and $C$, clearly shown from the non-overlapping CIs of the estimates among different correlations. Copula modeling enabled separate estimation of the correlation without assuming identical functional forms for the marginal distributions, as demonstrated by the accurate estimates under $T \sim \text{Weibull}$ and $C \sim \text{log-normal}$.

The results of correlation estimation in multiple runs of $(T, C)$ data generation and their bootstrap samples are presented in Table~\ref{tab:table2}. The mean estimate and MAE demonstrate the proposed method's accuracy in estimating the underlying correlation, and the CPs of the bootstrap 95\% CIs are near the desired 95\% level. Given that zero correlation implies independence under the normal (Gaussian) copula, the use of the normal copula in our simulations justifies the conclusion of independence between $T$ and $C$ when the bootstrap 95\% CI for the estimated correlation includes zero~\cite{Nelsen2006}.

\subsection{Comparison with the results of MLE}

The objective here is to compare the performance of our proposed method to that of conventional MLE. MLE constructs the likelihood function of $(X, \delta)$ by considering the (conditional CDF of $C$)$\times$(PDF of $T$) and (conditional CDF of $T$)$\times$(PDF of $C$) terms separately, depending on whether $\delta = 1$ or 0, and creates the composite likelihood by multiplying the terms. Specifically, the likelihood function is constructed as~\cite{Sorrell2023}
\begin{multline}
L(\Theta) = \prod_{i=1}^{n} \left\{\Pr(T_{i}=x_{i},\, C_{i}>x_{i})\right\}^{\delta_{i}} \cdot \left\{\Pr(C_{i}=x_{i},\, T_{i}>x_{i})\right\}^{1-\delta_{i}} \\
= \prod_{i=1}^{n}\bigg[\left\{\frac{\partial}{\partial S_{T}(x_{i};\theta_{T})}\Ccopula\!\left(S_{T}(x_{i};\theta_{T}),\, S_{C}(x_{i};\theta_{C})\right) \cdot f_{T}(x_{i};\theta_{T})\right\}^{\delta_{i}} \\
\times \left\{\frac{\partial}{\partial S_{C}(x_{i};\theta_{C})}\Ccopula\!\left(S_{T}(x_{i};\theta_{T}),\, S_{C}(x_{i};\theta_{C})\right) \cdot f_{C}(x_{i};\theta_{C})\right\}^{1-\delta_{i}}\bigg],
\end{multline}
where the parameter vector $\Theta = (\rho, \theta_{T}, \theta_{C})^{T}$, $\rho$: copula correlation parameter, $\theta_{T}, \theta_{C}$: marginal distribution parameters of $T, C$, respectively, $X_{i} = \min(T_{i}, C_{i}) = x_{i}$, $\delta_{i} = \mathbb{I}(T_{i} \leq C_{i})$: the observed time-to-event and event status of the $i$th subject, $\Ccopula(u, v)$: a copula function with uniform marginals and correlation parameter $\rho$, $S_{T}(x_{i};\theta_{T}), S_{C}(x_{i};\theta_{C})$: the marginal survival functions of $T, C$, respectively, and $f_{T}(x_{i};\theta_{T}), f_{C}(x_{i};\theta_{C})$: the marginal PDFs of $T, C$, respectively.

Subsequent numerical iterations to maximize the log-likelihood, usually by gradient descent, estimates the correlation parameter $\rho$ and the marginal distribution parameters $\theta_{T}$, $\theta_{C}$ simultaneously. We use the R code provided by Sorrell et al.~\cite{Sorrell2023} for MLE of the correlation parameter $\rho$ where the copulas and true correlations are varied as below:

The copula functions are varied among the normal, Clayton, Frank, and Gumbel copulas to estimate correlations of 0 (independence), 0.3, 0.5, and 0.8. The marginal distributions of $T$ and $C$ are set to follow a Weibull distribution.

For the copula dependence parameter values corresponding to a Kendall's tau of 0.3, 0.5, and 0.8, the normal copula's values are 0.4534, 0.7071, and 0.9511, the Clayton copula's 0.8571, 2, and 8, the Frank copula's 2.9174, 5.7363, and 18.1915, and the Gumbel copula's 1.4286, 2, and 5 (Supplementary Table~S1). The independence copula $u \cdot v$ is used to generate independent $(T, C)$ data. The marginal distribution parameters for Weibull (shape, scale) are $T \sim \text{Weibull}(0.63, 0.06)$ and $C \sim \text{Weibull}(0.86, 0.04)$, as noted previously.

\begin{table}[H]
\centering
\caption{Simulation results of correlation (dependence) estimation with the proposed method compared to those with maximum likelihood estimation (MLE) in dependently censored survival data $(T, C)^{\text{a}}$ and its bootstrap samples, where the two marginal survival times are Weibull distributed}
\label{tab:table3}
\begin{threeparttable}
\footnotesize
\setlength{\tabcolsep}{3.5pt}
\begin{tabular}{llccccccc}
\toprule
 & & \multicolumn{3}{c}{Proposed method} & & \multicolumn{3}{c}{MLE} \\
\cmidrule{3-5}\cmidrule{7-9}
Copula & True $\tau$ & Point Est. & Bootstr.$^{\text{b}}$ SE & Bootstr.\ 95\% CI & & Point Est. & Bootstr.\ SE & Bootstr.\ 95\% CI \\
\midrule
Indep. & 0 & 0.005 & 0.077 & $(-0.067,\; 0.186)$ & & 0.268 & 0.280 & $(0.096,\; 0.891)$ \\
\midrule
\multirow{3}{*}{Normal} & 0.3 & 0.249 & 0.085 & $(0.127,\; 0.471)$ & & 0.610 & 0.197 & $(0.096,\; 0.749)$ \\
 & 0.5 & 0.450 & 0.085 & $(0.347,\; 0.679)$ & & 0.654 & 0.202 & $(0.165,\; 0.815)$ \\
 & 0.8 & 0.720 & 0.033 & $(0.694,\; 0.824)$ & & 0.891 & 0.205 & $(0.289,\; 0.891)$ \\
\midrule
\multirow{3}{*}{Clayton} & 0.3 & 0.277 & 0.037 & $(0.265,\; 0.399)$ & & 0.373 & 0.078 & $(0.180,\; 0.473)$ \\
 & 0.5 & 0.549 & 0.040 & $(0.505,\; 0.652)$ & & 0.597 & 0.062 & $(0.388,\; 0.651)$ \\
 & 0.8 & 0.832 & 0.016 & $(0.700,\; 0.891)$ & & 0.921 & 0.093 & $(0.638,\; 0.955)$ \\
\midrule
\multirow{3}{*}{Frank} & 0.3 & 0.452 & 0.078 & $(0.247,\; 0.538)$ & & 0.498 & 0.227 & $(0.070,\; 0.909)$ \\
 & 0.5 & 0.554 & 0.075 & $(0.299,\; 0.557)$ & & 0.546 & 0.165 & $(0.340,\; 0.918)$ \\
 & 0.8 & 0.728 & 0.053 & $(0.613,\; 0.790)$ & & 0.669 & 0.191 & $(0.279,\; 0.963)$ \\
\midrule
\multirow{3}{*}{Gumbel} & 0.3 & 0.340 & 0.083 & $(0.247,\; 0.572)$ & & 0.013 & 0.139 & $(0.043,\; 0.641)$ \\
 & 0.5 & 0.479 & 0.076 & $(0.284,\; 0.568)$ & & 0.547 & 0.162 & $(0.076,\; 0.935)$ \\
 & 0.8 & 0.749 & 0.032 & $(0.602,\; 0.853)$ & & 0.910 & 0.163 & $(0.535,\; 0.972)$ \\
\bottomrule
\end{tabular}
\begin{tablenotes}[flushleft]\footnotesize
\item Abbreviations: Bootstr., bootstrap; CI, confidence interval; Est., estimate; Indep., independence copula; SE, standard error.
\item[$^{\text{a}}$] Sample sizes were 100, 200 or 500 for Kendall's tau of 0 or 0.8, and 1000 or 2000 for Kendall's tau of 0.3 or 0.5.
\item[$^{\text{b}}$] 200 bootstrap samples were used.
\end{tablenotes}
\end{threeparttable}
\end{table}

Table~\ref{tab:table3} shows the results of our proposed method and those of MLE in estimating the correlation between $T$ and $C$ when the copulas are varied and the marginals are Weibull-distributed. The point estimates, SEs, and 95\% CIs of the proposed method demonstrate robust performance across different copulas, regardless of their functional forms. The proposed method's ability to distinguish between weak or strong correlations is shown by the non-overlapping 95\% CIs, especially in the case of the Clayton copula. In contrast, the correlation estimates by MLE had largely biased point estimates, large SEs, and wide 95\% CIs. The consistently large SEs and resulting wide CIs indicate the instability of MLE in jointly estimating the copula and marginal distribution parameters. The MLE results were relatively better under the Clayton copula, although our proposed method showed smaller SEs and narrower 95\% CIs in this case as well.

Additional simulation results are provided as Supplementary material, where the proposed method is further utilized to accurately estimate the effect of a treatment in a simulated RCT.

\section{Real-Data Application: AIDS Clinical Trials Group (ACTG) Study 175}

\subsection{Data description and preparation}

A real-world dataset of a double-blind RCT among adults infected with the human immunodeficiency virus (HIV) whose CD4 T-cell counts were 200--500/mm$^{3}$ was obtained from the \textit{speff2trial} package in R~\cite{Juraska2022}. The study objective was to compare the efficacy of monotherapy with either zidovudine (also known as AZT) or didanosine vs.\ the combination therapies of AZT plus didanosine or AZT plus zalcitabine, resulting in a total of four treatment arms. The primary endpoint of the study was $\geq$50\% decline in CD4 T-cell count, progression of HIV to AIDS, or all-cause death, whichever came first. Early patient withdrawal due to deteriorating health or toxic effects of the drug occurred during the trial, which is strongly indicative of dependent censoring that is positively correlated with the primary endpoint. Therefore, the estimation of correlation between the primary endpoint and the competing event of patient withdrawal is necessary to unbiasedly estimate the efficacy of the treatment arms. The ACTG 175 data has been analyzed by several previous studies~\cite{Huang2008,Chen2010,Deresa2021} that also focused on the possible correlation between the time to primary endpoint and time to patient withdrawal.

Among the four treatment arms initially included in the data, only the two treatment arms of AZT alone ($N=532$) and AZT plus didanosine ($N=522$) were considered for a total of 1,054 patients. Among the 1,054 patients, 284 patients experienced the primary endpoint of a decline in CD4 T-cells, progression to AIDS, or death, while 381 patients withdrew from the trial and 389 were administratively censored at the end of study. The dependent censoring of patient withdrawal and administrative end-of-study censoring were combined into one competing event (against that of the primary endpoint) as $381+389 = 770$ patients censored. Since the study was a double-blind RCT with randomized treatment allocation, we expected the eight clinically relevant covariates of age, gender, race, intravenous drug use, hemophilia, baseline CD4 T-cell count, prior antiretroviral history, and disease symptoms indicator to be well-balanced between the two treatment arms. Either ANOVA or the Chi-squared test were used to test sufficient balance by treatment arms for the covariates at a significance level of 0.05.

Regarding the possible correlation between the two outcomes of time to the primary endpoint and time to either withdrawal or end-of-study censoring, three correlation scenarios were considered: a) estimation of correlation with the proposed method, b) assumed independence, and c) assumed correlation of Kendall's tau = 0.8. After either estimating or assuming the correlation between the survival endpoints, the marginal survival probability over time and the regression coefficient of the treatment arms variable in a univariable Cox regression model were estimated according to each correlation scenario. The marginal survival probability over time was plotted using the original ACTG 175 dataset, while the beta coefficient of the treatment arms variable was estimated in the original ACTG 175 dataset as well as in its 200 bootstrap samples for additional bootstrap SE and P-value calculations. The Wald statistic with bootstrap SEs were used for the bootstrap P-values. The \texttt{CG.Clayton()} function in the \textit{compound.Cox} R package~\cite{Emura2019} was used for marginal survival curve plotting, and the \texttt{dependCox.reg()} function for univariable Cox regression with dependent censoring. The conventional \texttt{coxph()} function in the \textit{survival} R package was used for the analysis scenario of assumed independence.

\subsection{Results of applying the proposed method}

Baseline characteristics of the ACTG Study 175 dataset ($N=1{,}054$) by the two treatment arms are shown in Supplementary Table~S3. A similar number of patients were allocated to each treatment (532 for monotherapy, 522 for combination therapy), and as expected from a double-blind RCT, the P-values showed that all relevant covariates were well-balanced between the treatment arms. The primary endpoint incidence and mean follow-up time differed substantially between the two treatment arms (both $P < 0.001$), with patients receiving the combination treatment showing a lower incidence of the primary endpoint and a longer mean follow-up.

\begin{figure}[H]
\centering
\includegraphics[width=\textwidth]{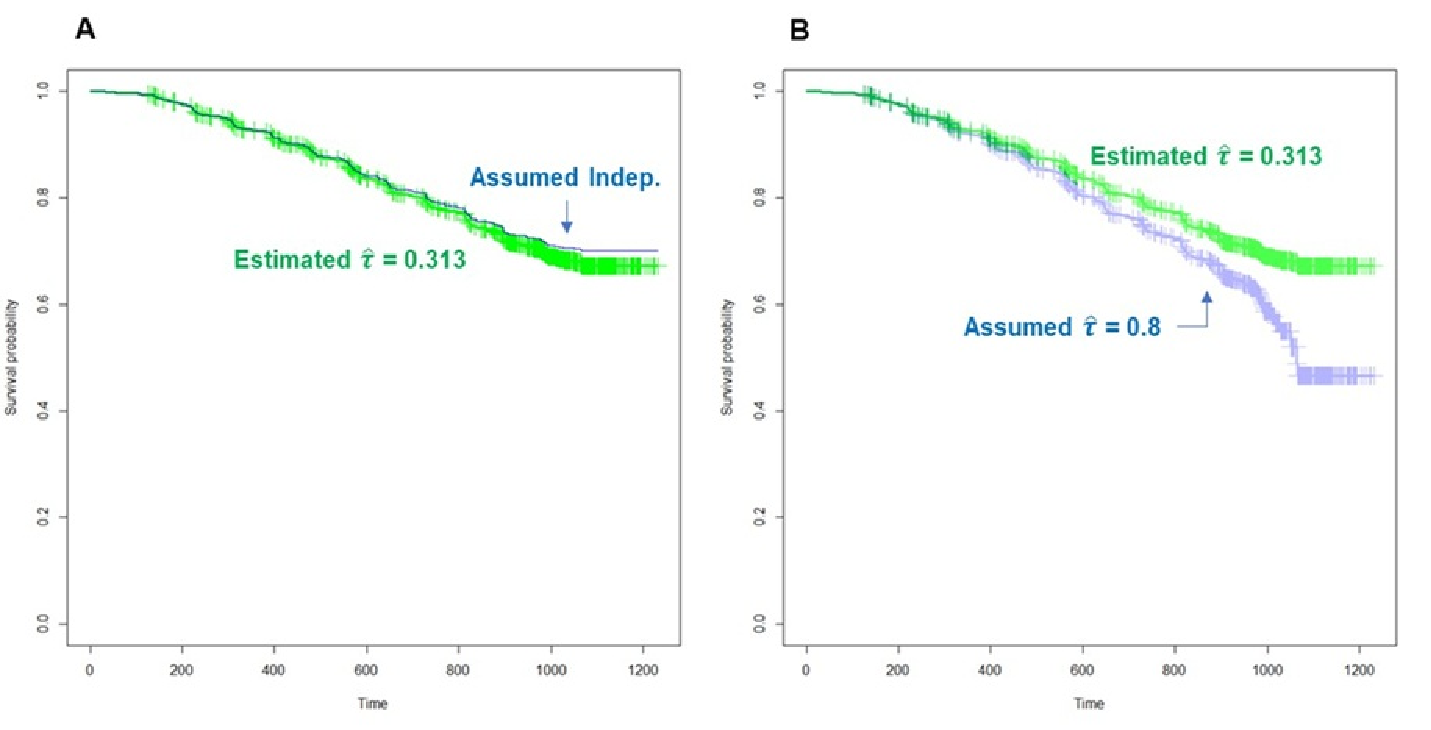}
\caption{Marginal survival curves of the time to the primary endpoint in the ACTG 175 dataset ($N=1{,}054$) under the proposed method's estimated correlation \textit{(in green)}, under independence \textit{(left, in blue; Part A)}, and under an assumed correlation of 0.8 \textit{(right, in blue; Part B)}, where the copula linking the time to the primary endpoint and the time to other endpoints (withdrawal from the trial or end of study censoring) is the Clayton copula.\\
{\footnotesize Abbreviations: ACTG, AIDS Clinical Trials Group Study; Indep., Independence ($\widehat{\tau}$ = 0)}}
\label{fig:figure1}
\end{figure}

Figure~\ref{fig:figure1} displays the marginal survival curves of the primary endpoint by the three estimated or assumed correlation scenarios. To plot the survival curve using our proposed correlation estimation method, we used the mean estimate from 200 bootstrap samples of the original ACTG 175 dataset. Compared to the survival curve based on our estimated Kendall's tau ($\widehat{\tau} = 0.313$, shown in green), the curve assuming independence between the survival endpoints overestimates the marginal survival of the primary endpoint (Part A, blue), while the curve assuming Kendall's tau $= 0.8$ underestimates it (Part B, blue). Accurate estimation of the marginal survival probability over time is essential for describing or predicting patient outcomes under a counterfactual scenario without early withdrawal or dependent censoring, thereby enabling a more objective comparison between the two treatments. In this context, the notably different survival curve under the assumed correlation of 0.8 (Part B, blue) illustrates the extent to which estimated marginal survival probabilities can vary with the degree of dependence between survival endpoints.

\begin{table}[H]
\centering
\caption{Results of correlation estimation with the proposed method vs.\ assumed correlations in subsequent regression coefficient estimation for a univariable Cox regression model of mono vs.\ combination treatment arms on the time to the primary endpoint in the ACTG 175 dataset ($N=1{,}054$), where the copula linking the time to the primary endpoint and the time to other endpoints (patient withdrawal or end of study censoring) is the Clayton copula}
\label{tab:table4}
\begin{threeparttable}
\footnotesize
\setlength{\tabcolsep}{2.5pt}
\begin{tabular}{lccccccccc}
\toprule
 & \multicolumn{3}{c}{Proposed method} & \multicolumn{3}{c}{Cause-specific hazards} & \multicolumn{3}{c}{Assumed correlation} \\
 & \multicolumn{3}{c}{of correlation est.} & \multicolumn{3}{c}{(Independence, $\widehat{\tau}=0$)} & \multicolumn{3}{c}{($\widehat{\tau}=0.8$)} \\
\cmidrule(lr){2-4}\cmidrule(lr){5-7}\cmidrule(lr){8-10}
Params. & Mean$^{\text{a}}$ Est. & Boot.$^{\text{b}}$ SE & Boot.\ $P^{\text{c}}$ & Mean Est. & Boot.\ SE & Boot.\ $P$ & Mean Est. & Boot.\ SE & Boot.\ $P$ \\
\midrule
$\beta_{T}$ & $-0.688$ & 0.118 & $<\!0.001$ & $-0.701$ & 0.116 & $<\!0.001$ & $-0.376$ & 0.085 & $<\!0.001$ \\
$\tau$ & 0.313 & 0.073 & $<\!0.001$ & 0 & -- & -- & 0.8 & -- & -- \\
\midrule
Params. & Orig.$^{\text{d}}$ Est. & Orig.\ SE & Orig.\ $P$ & Orig.\ Est. & Orig.\ SE & Orig.\ $P$ & Orig.\ Est. & Orig.\ SE & Orig.\ $P$ \\
\midrule
$\beta_{T}$ & $-0.675$ & 0.122 & $<\!0.001$ & $-0.704$ & 0.123 & $<\!0.001$ & $-0.372$ & 0.082 & $<\!0.001$ \\
$\tau$ & 0.370 & -- & -- & 0 & -- & -- & 0.8 & -- & -- \\
\bottomrule
\end{tabular}
\begin{tablenotes}[flushleft]\footnotesize
\item Abbreviations: ACTG, AIDS Clinical Trials Group Study; Boot., bootstrap; Est., estimate; Orig., original data; Params., parameters for estimation; SE, standard error.
\item[$^{\text{a}}$] The mean value of 200 regression coefficients estimated from the 200 bootstrap samples of the original ACTG 175 dataset.
\item[$^{\text{b}}$] The standard error (empirical standard deviation) of 200 regression coefficients estimated from the 200 bootstrap samples.
\item[$^{\text{c}}$] A two-sided P-value calculated from the Wald statistic $= \widehat{\beta}_{T} / \text{SE}(\widehat{\beta}_{T})$.
\item[$^{\text{d}}$] The single regression coefficient, SE, and P-value estimated from the original ACTG 175 dataset.
\end{tablenotes}
\end{threeparttable}
\end{table}

Table~\ref{tab:table4} compares the estimated regression coefficients of the combination treatment (vs.\ monotherapy) for the time to the primary endpoint in a univariable Cox regression model among the three scenarios of estimated or assumed correlations between the primary endpoint and other endpoints (patient withdrawal or end of study censoring). Also in separate rows are the mean estimate of 200 bootstrap samples of the original ACTG 175 dataset and the single estimate of the original dataset itself. As expected, the estimated regression coefficients, SEs, and P-values of the original dataset and those of its bootstrap samples are similar.

Comparing the three estimated or assumed correlations column-wise, the ordering of the relative sizes of the estimated regression coefficients shows that the effect estimate is largest under assumed independence between the survival endpoints ($-0.701$), slightly smaller under the estimated correlation of 0.313 by the proposed method ($-0.688$), and smallest under an assumed correlation of 0.8 ($-0.376$). This is in agreement with the previous studies by Huang and Zhang and Chen~\cite{Huang2008,Chen2010}, while difficult to directly compare with Deresa and Van Keilegom~\cite{Deresa2021} due to their use of a linear regression model. Overall, the combination treatment is clearly superior over monotherapy in terms of the primary endpoint with protective HRs of $\exp(-0.688) = 0.50$, $\exp(-0.701) = 0.49$, or $\exp(-0.376) = 0.69$ across all three correlation scenarios, which is also in agreement with the studies above.

A point of note is the difference in the estimated correlation between our study and Deresa and Van Keilegom~\cite{Deresa2021}. First, we estimated a mean correlation of 0.313 between the time to primary endpoint and time to other endpoints (patient withdrawal or end of study censoring), while the previous study's correlation estimation of 0.458 ($= 2/\pi \cdot \arcsin(0.659)$) was between the time to primary endpoint and time to patient withdrawal, treating end of study censoring as administrative independent censoring. This is clearly reasonable, while we believe that combining all other events into one dependent censoring event to estimate its correlation with the event of interest is also a reasonable approach. Second, since the ACTG study 175 was a double-blind RCT, we confirmed that all eight covariates were well-balanced between the two treatment arms and proceeded with a univariable Cox regression of the treatment arm variable upon the time to primary event. The previous studies differ from ours in additionally adjusting for these covariates, which led to different beta coefficient estimates. However, the relative effect sizes by the estimated correlation, assumed independence, or assumed correlation of 0.8 are in agreement among all studies including ours. Conclusively, the better efficacy of combination therapy (vs.\ monotherapy) under the explicit estimation of correlation between the survival endpoints is not as large compared to that of assumed independence, but larger than that under the possibly incorrect correlation of 0.8.

\section{Discussion}

The current study proposed a novel method to estimate the correlation and marginal distributions in dependently censored survival data, where only the minimum of the two time-to-events is observable. Assuming the marginal survival times follow exponential, Weibull, or log-normal distributions, a copula correlation parameter is identified using method-of-moments bagging combined with a gradient descent local search. Unlike previous studies, the proposed method does not require informative covariates to estimate the correlation between survival times. A simulation study demonstrated accurate and efficient estimations in terms of the MAE, CP, and bootstrap CIs (Tables~\ref{tab:table1}--\ref{tab:table3}). Real-world applicability was examined with an RCT dataset of HIV patients, confirming the method's usefulness in handling correlated survival outcomes when estimating treatment efficacy (Table~\ref{tab:table4}).

A strength of this study is the estimation of the correlation between survival endpoints under the relatively general assumption that the marginal survival times follow commonly used parametric distributions (Tables~\ref{tab:table1} and~\ref{tab:table2}). Notably, the proposed method does not rely on informative covariates linking the two survival times. Unlike earlier studies that assumed the copula correlation parameter to be known, Emura and Chen~\cite{Emura2016} did estimate the correlation parameter $\rho$ using cross-validation within a survival prediction framework. Their approach was to choose $\rho$ that maximizes the cross-validated Harrell's c-index, under the rationale that the value resulting in the best prediction of actual survival times would be its true value. However, this approach relies on the existence of highly predictive covariates. The recent work by Jo et al.~\cite{Jo2023} is similar in that an informative covariate $P$ (= time to progression of disease) correlated with both $T$ (= survival time of pancreatic cancer patients) and $C$ (= dependent censoring time) is assumed to exist.

Another strength of the proposed method is its good performance compared to MLE when the marginals are bivariate Weibull-distributed (Table~\ref{tab:table3}). Several previous studies have demonstrated that simultaneous estimation of copula correlation and marginal distribution parameters using MLE produces biased estimates with large standard errors~\cite{Schneider2023,Emura2016,Chen2010,Michimae2022}, which was confirmed in the current study. The study by Czado and Van Keilegom~\cite{Czado2023} extended parametric identifiability from the bivariate normal distribution and the normal (Gaussian) copula to other marginal distributions and copulas, such as the log-normal and Weibull marginal distributions and the Clayton, Frank, and Gumbel copulas. Their theorems state that dependently censored survival data with the parametric distributions and copulas noted above is identifiable from the usual likelihood construction and subsequent MLE. However, we empirically verified the instability of MLE in the case of bivariate Weibull marginals linked through normal, Clayton, Frank, or Gumbel copulas (Table~\ref{tab:table3}), and the apparent disagreement between their statements and our results may require further study.

Additionally, the accuracy of our method indicated by the MAEs in Tables~\ref{tab:table1} and~\ref{tab:table2} deserves further discussion. Although an MAE of approximately 0.05 may not appear highly accurate, this result should be interpreted in the context of the previous ``non-identifiability'' problem. Earlier copula-based approaches either assumed that the true correlation was known, or relied heavily on sensitivity analyses to approximate its plausible range. Given that only the minimum of the two survival times ($T$ or $C$) is observed, the feasibility of obtaining a direct point estimate for the correlation may be considered notable in itself. Furthermore, consistency between the MAEs and their corresponding standard errors provides additional evidence supporting the robustness of our estimation method.

Several limitations of this study should also be acknowledged. First, we did not separately consider end-of-study administrative censoring, which is expected to be independent of the event time $T$. In this case, further correlation estimation wouldn't be needed and the usual cause-specific hazards approach would suffice. Instead, we combined all possible time-to-events other than $T$, including end-of-study censoring, into a single dependent censoring time $C$. Second, although GMM estimation is known to be versatile, it also has drawbacks such as its non-uniqueness of solutions~\cite{Casella2024}. A two-stage approach was thus employed to ensure accurate estimation of a single, unique correlation: we first identified the plausible range of the correlation parameter using method-of-moments bagging, followed by a point estimation within this range using gradient descent. Third, the proposed method currently requires larger sample sizes to accurately estimate low to moderate correlations (Kendall's tau of 0.3 or 0.5) compared to zero or high correlations (Kendall's tau of 0 or 0.8). Sample sizes of approximately 1,000 to 2,000 are needed to reliably distinguish correlations of 0.3 or 0.5, whereas smaller samples of 100 to 500 are sufficient to estimate correlations at or near the extremes (0 or 0.8). This aligns intuitively with the fact that zero or very high correlation typically result in more distinctive data patterns than intermediate correlations. Data augmentation may be considered to overcome this limitation, and we are currently exploring the use of generative artificial intelligence algorithms to effectively augment such datasets.

In summary, the current study utilized copula dependence modeling to parametrically identify the correlation in dependently censored survival data under exponential, Weibull, or log-normally distributed survival times. The proposed method estimates the copula correlation parameter and marginal survival distributions in conjunction, which subsequently enables accurate estimation of marginal treatment effects. Our method has potential biomedical applications where the marginal survival or hazard is of interest, such as in RCTs aiming to estimate the effect of a drug on patient survival without the influence of dependent censoring.

\subsection*{Full list of abbreviations}

\noindent
ACTG: AIDS clinical trials group;
AIDS: acquired immunodeficiency syndrome;
ANOVA: analysis of variance;
AZT: zidovudine;
Bagging: bootstrap aggregating;
CD4 T-cell: T helper cells (CD: cluster of differentiation);
CDF: cumulative distribution function;
CG: copula-graphic;
CI: confidence interval;
CP: coverage probability;
GMM: generalized method-of-moments;
HIV: human immunodeficiency virus;
HR: hazard ratio;
K-M: Kaplan-Meier;
MAE: mean absolute error;
MLE: maximum likelihood estimation;
OS: overall survival;
PDF: probability density function;
PH: proportional hazards;
RCT: randomized clinical trial;
SE: standard error

\clearpage
\appendix
\renewcommand{\thesection}{S\arabic{section}}
\renewcommand{\thetable}{S\arabic{table}}
\renewcommand{\thefigure}{S\arabic{figure}}
\setcounter{section}{0}
\setcounter{table}{0}
\setcounter{figure}{0}

\begin{center}
\Large\textbf{Supplementary Material for:}\\[6pt]
\large\textbf{Bootstrap-Aggregated Method of Moments Estimation of the Copula Correlation Parameter for Marginal Survival Inference under Dependent Censoring}
\end{center}

\section{Supplementary Methods S1 for ``2.1 Setup and notation''}

\textbf{Supplementary Table S1: Functional forms and corresponding Kendall's tau for several parametric copulas}

\begin{table}[H]
\centering
\small
\begin{tabular}{lll}
\toprule
Copula type & Copula function with parameter $\theta$ & Corresponding Kendall's tau \\
\midrule
Normal & $\displaystyle\frac{1}{2\pi\sqrt{1-\theta^{2}}}\int_{-\infty}^{\Phi_{T}^{-1}(u)}\!\int_{-\infty}^{\Phi_{C}^{-1}(v)}\!\exp\!\left[-\frac{t^{2}-2\theta tc+c^{2}}{2(1-\theta^{2})}\right]dc\,dt$ & $\displaystyle\frac{2}{\pi}\arcsin(\theta)$ \\[18pt]
Clayton & $\displaystyle\left(u^{-\theta}+v^{-\theta}-1\right)^{-1/\theta}$ & $\displaystyle\frac{\theta}{2+\theta}$ \\[12pt]
Gumbel & $\displaystyle\exp\!\left[-\left\{(-\ln u)^{\theta}+(-\ln v)^{\theta}\right\}^{1/\theta}\right]$ & $\displaystyle\frac{\theta-1}{\theta}$ \\[12pt]
Frank & $\displaystyle-\frac{1}{\theta}\ln\!\left[1+\frac{(e^{-\theta u}-1)(e^{-\theta v}-1)}{e^{-\theta}-1}\right]$ & $\displaystyle 1-\frac{4}{\theta}\left\{1-\frac{1}{\theta}\int_{0}^{\theta}\frac{t}{e^{t}-1}\,dt\right\}$ \\[12pt]
\bottomrule
\end{tabular}
\end{table}

Survival time distributions of the exponential, Weibull, or log-normal marginals as special cases of the generalized gamma density are expressed as follows. The generalized gamma distribution~\cite{Stacy1965} is a generalization of the gamma distribution that nests several other distributions widely used in parametric models of survival times: namely, the exponential, Weibull, and log-normal distributions (as well as the gamma distribution itself). For the generalized gamma PDF~\cite{Collett2023}
\begin{equation}
f(x) = \frac{\alpha\lambda^{\gamma\alpha}x^{\gamma\alpha-1}\exp\{-(\lambda x)^{\alpha}\}}{\Gamma(\gamma)},\quad \lambda:\text{scale},\;\gamma:\text{shape},\;\alpha:\text{add.\ shape},\;\text{all}>0,\;\text{for}\;0\leq x<\infty,
\end{equation}

\begin{enumerate}[label=\roman*)]
\item $\alpha = \gamma = 1$ gives the exponential distribution PDF $f(x) = \lambda e^{-\lambda x}$ with scale $\lambda$,
\item $\gamma = 1$ the Weibull PDF $f(x) = \alpha\lambda^{\alpha}x^{\alpha-1}\exp\{-(\lambda x)^{\alpha}\}$ with shape $\alpha$ and scale $\lambda^{\alpha}$,
\item $\gamma \to \infty$ the log-normal PDF $f(x) = \frac{1}{x\sigma\sqrt{2\pi}}\exp\left[-\frac{1}{2}\frac{\{\log(x)-\mu\}^{2}}{\sigma^{2}}\right]$,
\end{enumerate}
with mean $\mu = -\log(\lambda) + \frac{\log(\gamma)}{\alpha}$ and variance $\sigma^{2} = \frac{1}{\gamma\alpha^{2}}$ for $\log(X)$.

Parametric marginal survival times $T$ and $C$ were generated by using the inverse survival functions of these three distributions from the generalized gamma family.

\section{Supplementary Methods S2 for ``2.3 Moment functions and generalized method-of-moments (GMM) estimation''}

For the bivariate normal random vector
\begin{equation}
(Y_{1}, Y_{2})' \sim N_{2}\!\left(\begin{bmatrix}\mu_{T}\\\mu_{C}\end{bmatrix},\begin{bmatrix}\sigma_{T}^{2} & \rho\sigma_{T}\sigma_{C}\\\rho\sigma_{T}\sigma_{C} & \sigma_{C}^{2}\end{bmatrix}\right),
\end{equation}
let $D = Y_{1}-Y_{2} \sim N(m, s^{2})$, $m = \mu_{T}-\mu_{C}$, $s^{2}=\sigma_{T}^{2}+\sigma_{C}^{2}-2\rho\sigma_{T}\sigma_{C}$, and
\begin{equation}
\kappa = -\frac{m}{s},\qquad \zeta_{1} = \frac{\varphi(\kappa)}{\Phi(\kappa)},\qquad \zeta_{2} = \frac{\varphi(\kappa)}{1-\Phi(\kappa)} = \frac{\varphi(\kappa)}{\Phi(-\kappa)},
\end{equation}
where $\varphi$ and $\Phi$ are the standard normal PDF and CDF.

Two mathematical facts~\cite{Hogg2013} regarding the conditioning of jointly normal $(Y_{1}, D)$, $(Y_{2}, D)$, and the one-sided truncation of $D$ are first stated as

\noindent\textbf{(F1)} Conditioning: For $(Y_{1}, D)$ with $\text{Cov}(Y_{1}, D) = c_{1}$ and $(Y_{2}, D)$ with $\text{Cov}(Y_{2}, D) = c_{2}$,
\begin{align}
E[Y_{1} \mid D=d] &= \mu_{T} + \frac{c_{1}}{s^{2}}(d-m),\qquad \text{Var}[Y_{1} \mid D=d] = \sigma_{T}^{2} - \frac{c_{1}^{2}}{s^{2}}, \\
E[Y_{2} \mid D=d] &= \mu_{C} + \frac{c_{2}}{s^{2}}(d-m),\qquad \text{Var}[Y_{2} \mid D=d] = \sigma_{C}^{2} - \frac{c_{2}^{2}}{s^{2}}.
\end{align}

\noindent\textbf{(F2)} Truncation: For $D \sim N(m, s^{2})$,
\begin{align}
E[D \mid D \leq 0] &= m - s\zeta_{1},\qquad \text{Var}[D \mid D \leq 0] = s^{2}(1 - \kappa\zeta_{1} - \zeta_{1}^{2}), \\
E[D \mid D > 0] &= m + s\zeta_{2},\qquad \text{Var}[D \mid D > 0] = s^{2}(1 + \kappa\zeta_{2} - \zeta_{2}^{2}).
\end{align}

The theoretical moments in 2.3 are derived as

\begin{enumerate}[label=\roman*)]
\item $p(\theta) = \Pr(D \leq 0) = \Phi\!\left(\frac{0-m}{s}\right) = \Phi(\kappa)$.

\item $\mu_{1}(\theta) = E[Y_{1} \mid D \leq 0]$:

The covariance $c_{1} = \text{Cov}(Y_{1}, Y_{1}-Y_{2}) = \sigma_{T}^{2} - \text{Cov}(Y_{1}, Y_{2}) = \sigma_{T}(\sigma_{T}-\rho\sigma_{C})$. By the law of iterated expectation and (F1), (F2),
\begin{multline}
\mu_{1}(\theta) = E[Y_{1} \mid D \leq 0] = E[E[Y_{1} \mid D] \mid D \leq 0] \\
= \mu_{T} + \frac{c_{1}}{s^{2}}(E[D \mid D \leq 0] - m) = \mu_{T} - \frac{\sigma_{T}(\sigma_{T}-\rho\sigma_{C})}{s}\zeta_{1}.
\end{multline}

\item $\sigma_{1}^{2}(\theta) = \text{Var}[Y_{1} \mid D \leq 0]$:

By the law of total variance and (F1), (F2),
\begin{equation}
\sigma_{1}^{2}(\theta) = \left(\frac{c_{1}}{s^{2}}\right)^{2}\text{Var}[D \mid D \leq 0] + \left(\sigma_{T}^{2} - \frac{c_{1}^{2}}{s^{2}}\right) = \sigma_{T}^{2} - \left\{\frac{\sigma_{T}(\sigma_{T}-\rho\sigma_{C})}{s}\right\}^{2}(\kappa\zeta_{1}+\zeta_{1}^{2}).
\end{equation}

\item $\mu_{2}(\theta) = E[Y_{2} \mid D > 0]$:

The covariance $c_{2} = \text{Cov}(Y_{2}, Y_{1}-Y_{2}) = -\sigma_{C}(\sigma_{C}-\rho\sigma_{T})$. As in ii),
\begin{equation}
\mu_{2}(\theta) = E[Y_{2} \mid D > 0] = \mu_{C} - \frac{\sigma_{C}(\sigma_{C}-\rho\sigma_{T})}{s}\zeta_{2}.
\end{equation}

\item $\sigma_{2}^{2}(\theta) = \text{Var}[Y_{2} \mid Y_{2} < Y_{1}]$: As in iii),
\begin{equation}
\sigma_{2}^{2}(\theta) = \sigma_{C}^{2} - \left\{\frac{\sigma_{C}(\sigma_{C}-\rho\sigma_{T})}{s}\right\}^{2}(-\kappa\zeta_{2}+\zeta_{2}^{2}).\qquad\blacksquare
\end{equation}
\end{enumerate}

\section{Supplementary Methods S3 for ``2.4 Feasible parameter space using copula-graphic bounds''}

Additional details on defining the rectangular search region $\Pspace$ for $(\mu_{T}, \sigma_{T}^{2}, \mu_{C}, \sigma_{C}^{2})$ are provided here. First, the copula graphic (CG) estimator~\cite{Emura2018,Zheng1995,Zheng1996} of marginal survival at some time $x$ iterates through a bisection root-finding algorithm by relating $F(x)$ of $T$ and $G(x)$ of $C$ on the unit square. Using an Archimedean copula and its generator function $\varphi(\cdot)$, the CG estimator $\widehat{S}_{T}(x)$ is expressed in closed form as
\begin{equation}
\widehat{S}_{T}(x) = \varphi_{\theta}^{-1}\!\left[\sum_{\substack{x_{i}\leq x \\ \delta_{i}=1}} \varphi_{\theta}\!\left(\frac{n_{i}-1}{n}\right) - \varphi_{\theta}\!\left(\frac{n_{i}}{n}\right)\right],\quad 0 \leq x \leq \max_{i}(x_{i}),
\end{equation}
where $n_{i} = \sum_{l=1}^{n} I(x_{l} \geq x_{i})$ is the number of subjects at risk at event time $x_{i}$ and assuming no ties in event times.

The CG estimator of marginal survival functions is utilized to set lower and upper bounds on the four marginal distribution parameters, while the range of the correlation parameter in terms of Kendall's tau is divided into the four possible ranges of none: $(-0.1, 0.15]$, low: $(0.15, 0.4]$, moderate: $(0.4, 0.65]$, or high: $(0.65, 0.9]$. In the case of bivariate log-normal $(T, C)$, the following steps define $\Pspace$:

\begin{enumerate}[label=\roman*)]
\item Set the representative values of $\rho$ as 0 (none), 0.3 (low), 0.5 (moderate), or 0.8 (high) in terms of Kendall's tau.
\item Estimate four sets of marginal survival functions of $T$ and $C$ under each representative value of $\rho$ utilizing the CG estimator.
\item For each of the four sets of CG estimators of $T$ and $C$, fit parametric (log-normal) survival distributions, e.g.\ ordinary least squares, with initial parameters such as $\widehat{\mu}_{T} = \widehat{\mu}_{C} = 0$ and $\widehat{\sigma}_{T}^{2} = \widehat{\sigma}_{C}^{2} = 1$.
\item Repeat iii) a certain number of times, e.g.\ 200 iterations, to obtain stable mean estimates of the marginal distribution parameters for each of the four sets of CG estimators.
\item For the different estimates of each parameter under the four correlations of tau1 to tau4, obtain 1st and 3rd quartiles Q1 and Q3 to form lower and upper bounds for each of the marginal parameters as $Q1 - 1.5 \times (Q3 - Q1)$ and $Q3 + 1.5 \times (Q3 - Q1)$.
\end{enumerate}

\section{Supplementary Results S1: Estimation of marginal survival and the effect of a binary treatment variable}

\subsection{Data generation and simulation parameter settings}

In biomedical studies with dependent censoring, the primary objective is often to obtain unbiased estimates of the marginal survival probability (or hazard rate) for the event of interest. Researchers are also typically interested in assessing the effect (regression coefficient) of a treatment or exposure on this marginal survival or hazard. Accordingly, we simulated scenarios representing these subsequent analyses following the estimation of correlation under dependent censoring~\cite{Huang2008}.

500 samples from the Weibull and exponential marginal distributions of $T \sim \text{Weibull}(2, 0.25)$ and $C \sim \text{Exponential}(0.2)$, are linked through a Clayton copula with a simulated Kendall's tau $= 0.791$ (true tau $= 0.8$), and the marginal survival curves of $T$ are subsequently plotted under scenarios a) to d) as:
\begin{enumerate}[label=\alph*)]
\item using the proposed method to estimate the correlation between $T$ and $C$ as 0.769,
\item incorrectly assuming zero correlation or independence (i.e., cause-specific hazards),
\item and d) incorrectly specifying the correlation to be 0.3 and 0.9, respectively.
\end{enumerate}

This example is then applied to a hypothetical RCT setting with a single binary treatment variable, where the estimation of its efficacy (regression coefficient) is the study objective. In the RCT setting of evaluating a new treatment to cure a disease, the time to event of interest $T$ is the overall survival (OS) time of a patient, while $C$ is the dependently censored time of patient withdrawal from the trial owing to deteriorating health or adverse side-effects of the new treatment. Thus, a strongly positive correlation between $T$ and $C$ is expected. The new treatment variable `Trt' is generated as $\text{Trt} \sim \text{Bernoulli}(0.5)$ with equal probability of assignment to the new treatment or a placebo. As the treatment assignment is randomized, all other covariates such as age, gender, and disease characteristics are assumed to be well-balanced between the treatment and placebo groups such that no adjustment for covariates is needed.

A semi-parametric Cox proportional hazards (PH) model is then assumed as
\begin{equation}
h(x \mid \text{Trt}) = h_{0}(x) \cdot \exp[\beta' \cdot \text{Trt}],\qquad X = \min(T, C)
\end{equation}
with baseline hazards of $T$ and $C$ as $h_{0,T}(t) = \alpha_{T}\lambda_{T}t^{\alpha-1}$ and $h_{0,C}(c) = \lambda_{C}$, respectively. The hazard functions are thus set as
\begin{align}
h_{T}(t \mid \text{Trt}) &= 2 \cdot 0.25 \cdot t^{(2-1)} \cdot \exp(\beta_{T} \cdot \text{Trt}) = 0.5 \cdot t \cdot \exp(\beta_{T} \cdot \text{Trt}), \\
h_{C}(c \mid \text{Trt}) &= 0.2 \cdot \exp(\beta_{C} \cdot \text{Trt}),
\end{align}
and the correlated survival times $(T, C)$ are generated as
\begin{align}
T &= \left[-\frac{\log(u)}{\{0.25 \cdot \exp(\beta_{T} \cdot \text{Trt})\}}\right]^{1/2},\qquad
C = -\frac{\log(v)}{\{0.2 \cdot \exp(\beta_{C} \cdot \text{Trt})\}},
\end{align}
for $\Ccopula(u,v) = (u^{-\rho}+v^{-\rho}-1)^{-1/\rho}$ from a Clayton copula with correlation parameter $\rho = 8$, corresponding to a Kendall's tau $= \rho/(2+\rho) = 0.8$ (Supplementary Table~S1).

The true effect sizes or regression coefficients of the new treatment on $T$ and $C$, $\beta_{T}$ and $\beta_{C}$, are set as $-0.5$ and 0.2, respectively. These beta values correspond to a hazard ratio (HR) $= \exp(-0.5) = 0.61$ of the new treatment on $T$ (= OS time), and HR $= \exp(0.2) = 1.22$ of the new treatment on $C$ (= time to patient withdrawal or dependent censoring). That is, the new treatment is largely effective in prolonging patient survival, while having a slightly detrimental effect on, or increasing the hazard of, patient withdrawal from the trial. No other censoring or competing events were assumed, as one can combine all events other than the event of interest into a single dependent censoring event.

200 bootstrap samples are taken from the initially generated $(T, C)$, and the regression coefficients $\beta_{T}$ and $\beta_{C}$ of the treatment variable `Trt' are estimated under the correlation scenarios a) to d). The \texttt{dependCox.reg()} function from the \textit{compound.Cox} package in R~\cite{Emura2019} is used for scenarios a), c), and d), while the conventional \texttt{coxph()} function from the \textit{survival} package is used for b).

\subsection{Simulation results}

The main purpose of Supplementary Figures~S1 and~S2 is to visually demonstrate the unbiased marginal survival curve when the correlation between $T$ and $C$ is correctly estimated by our proposed method, compared to the biased marginal survivals under incorrectly specified correlations. The well-known Kaplan-Meier (K-M) estimator assumes independent censoring, corresponding to the over-estimated survival curve in the right of Figure~S1. For positively (or negatively) correlated time-to-events $T$ and $C$, the K-M curve will always over-estimate (or under-estimate) the marginal survival curve as shown. Naturally, if the correlation is wrongly assumed as smaller (or larger) than the actual positive correlation, this will result in over-estimation (or under-estimation) of the marginal survival probabilities (Figure~S2).

In the hypothetical RCT scenario of a new treatment to a disease, the time to event of interest $T$ may be the OS time of a patient, while the dependent censoring time $C$ may be the observed time until patient withdrawal from the trial due to deteriorating health or adverse side-effects of the new treatment. Bias in the marginal survival of $T$, especially in the K-M estimate, implies that the patients' marginal OS probability, and thus the efficacy of the new treatment, is incorrectly estimated.

\begin{figure}[H]
\centering
\includegraphics[width=\textwidth]{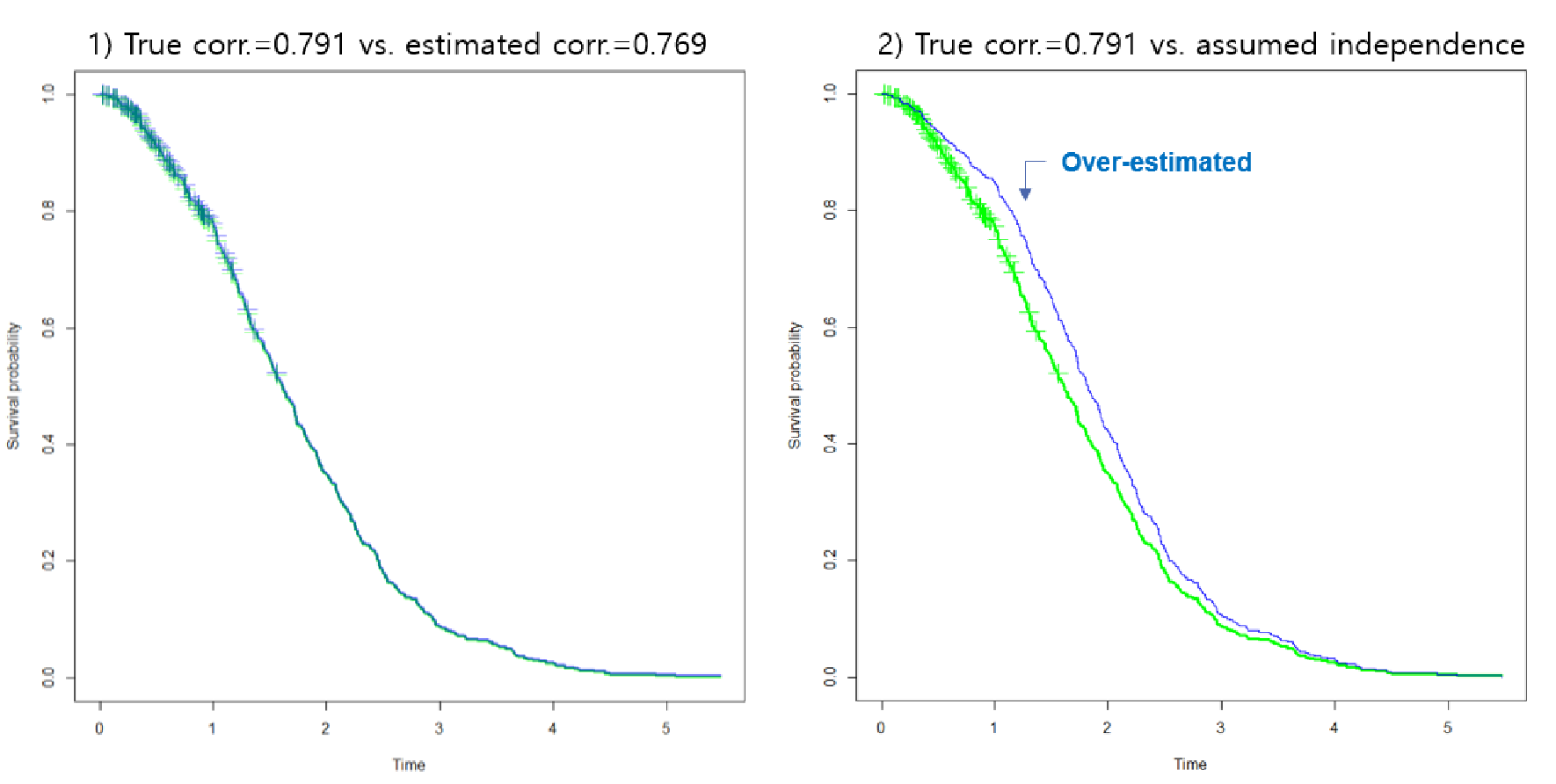}
\caption{Simulation results of marginal survival curves of the time to event of interest $T$ in dependently censored survival data $(T, C)^{\text{a}}$, where the ``true'' marginal survival curve of $T$ with Kendall's tau $= 0.791$ is plotted in green, and the marginal survival curves from either estimating the correlation between $(T, C)$ via the proposed method or wrongly assuming the correlation to be zero (independence) are plotted in blue.\\
{\footnotesize $^{\text{a}}$ The marginal distributions of $T$ and $C$ are Weibull and Exponentially distributed, and the copula linking the marginal distributions is the Clayton copula.}}
\label{fig:figureS1}
\end{figure}

\begin{figure}[H]
\centering
\includegraphics[width=\textwidth]{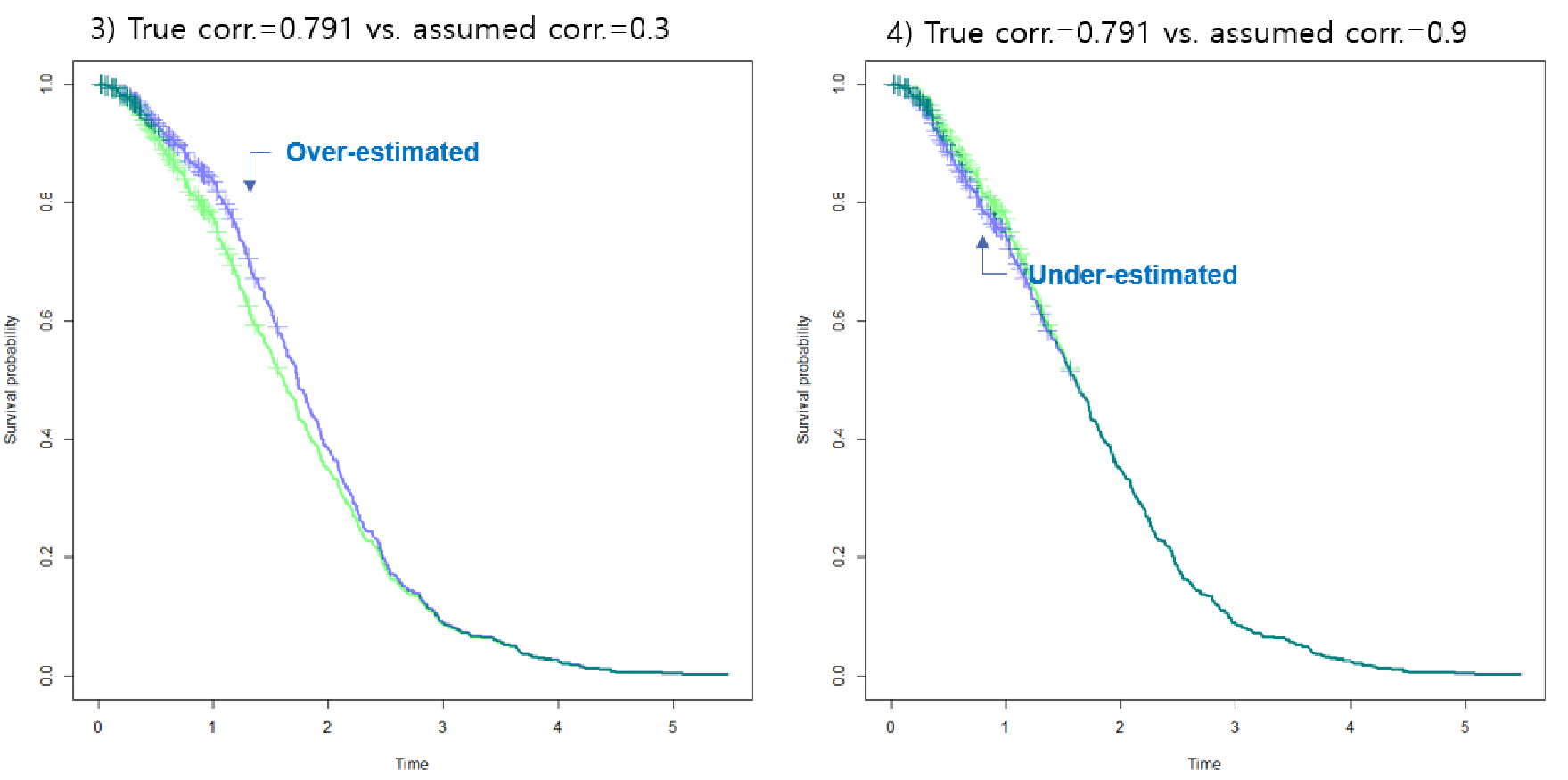}
\caption{Simulation results of marginal survival curves of the time to event of interest $T$ in dependently censored survival data $(T, C)^{\text{a}}$, where the ``true'' marginal survival curve of $T$ with Kendall's tau $= 0.791$ is plotted in green, and the marginal survival curves from wrongly assuming the correlation between $(T, C)$ to be either 0.3 or 0.9 are plotted in blue.\\
{\footnotesize $^{\text{a}}$ The marginal distributions of $T$ and $C$ are Weibull and Exponentially distributed, and the copula linking the marginal distributions is the Clayton copula.}}
\label{fig:figureS2}
\end{figure}

\begin{table}[H]
\centering
\caption{Simulation results of regression coefficients estimation for a univariable Cox regression model in dependently censored survival data $(T, C)^{\text{a}}$ with true Kendall's tau correlation of 0.8, under correlation estimation ($=0.769$) with the proposed method or wrongly assumed independence}
\label{tab:tableS2.1}
\begin{threeparttable}
\small
\begin{tabular}{lcccccc}
\toprule
 & & \multicolumn{2}{c}{Proposed method} & & \multicolumn{2}{c}{Cause-specific hazards} \\
 & & \multicolumn{2}{c}{of correlation est.} & & \multicolumn{2}{c}{(Independence, $\widehat{\tau}=0$)} \\
\cmidrule{3-4}\cmidrule{6-7}
Param. & True value & Mean est. / MPE & Bootstr. SE & & Mean est. / MPE & Bootstr. SE \\
\midrule
$\beta_{T}$ & $-0.5$ & $-0.482$ / 3.58\% & 0.096 & & $-0.718$ / $-43.6\%$ & 0.109 \\
$\beta_{C}$ & 0.2 & 0.187 / 6.58\% & 0.191 & & 0.545 / $-172\%$ & 0.197 \\
$\tau$ & 0.8 & 0.769 / 3.56\% & 0.055 & & 0.000 / $-100\%$ & -- \\
\bottomrule
\end{tabular}
\begin{tablenotes}[flushleft]\footnotesize
\item Abbreviations: Bootstr., bootstrap; Est., estimate; MPE, mean percentage error; Param., parameter for estimation; SE, standard error.
\item[$^{\text{a}}$] The marginal distributions of $T$ and $C$ are Weibull and Exponentially distributed, and the copula linking the marginal distributions is the Clayton copula.
\end{tablenotes}
\end{threeparttable}
\end{table}

\begin{table}[H]
\centering
\caption{Simulation results of regression coefficients estimation for a univariable Cox regression model in dependently censored survival data $(T, C)^{\text{a}}$ with true Kendall's tau correlation of 0.8, under wrongly assumed correlations of 0.3 and 0.9}
\label{tab:tableS2.2}
\begin{threeparttable}
\small
\begin{tabular}{lcccccc}
\toprule
 & & \multicolumn{2}{c}{Incorrectly assumed} & & \multicolumn{2}{c}{Incorrectly assumed} \\
 & & \multicolumn{2}{c}{correlation ($\widehat{\tau}=0.3$)} & & \multicolumn{2}{c}{correlation ($\widehat{\tau}=0.9$)} \\
\cmidrule{3-4}\cmidrule{6-7}
Param. & True value & Mean est. / MPE & Bootstr. SE & & Mean est. / MPE & Bootstr. SE \\
\midrule
$\beta_{T}$ & $-0.5$ & $-0.619$ / $-23.8\%$ & 0.105 & & $-0.436$ / 12.8\% & 0.088 \\
$\beta_{C}$ & 0.2 & 0.517 / $-159\%$ & 0.193 & & $-0.130$ / 164.9\% & 0.117 \\
$\tau$ & 0.8 & 0.300 / $-62.5\%$ & -- & & 0.900 / 12.5\% & -- \\
\bottomrule
\end{tabular}
\begin{tablenotes}[flushleft]\footnotesize
\item Abbreviations: Bootstr., bootstrap; Est., estimate; MPE, mean percentage error; Param., parameter for estimation; SE, standard error.
\item[$^{\text{a}}$] The marginal distributions of $T$ and $C$ are Weibull and Exponentially distributed, and the copula linking the marginal distributions is the Clayton copula.
\end{tablenotes}
\end{threeparttable}
\end{table}

Supplementary Tables~S2.1 and~S2.2 show the simulation results of estimating the regression coefficients in a Cox PH regression model of a binary treatment variable when $T$ and $C$ are dependent. Under our simulation settings, the average proportion of $T$ (or $C$) observed was 77.2\% (or 22.8\%).

First, the proposed method of correlation estimation performed well with a mean estimate of 0.769 and standard error of 0.055, compared to the simulated Kendall's tau of 0.791 (true Kendall's tau of 0.8). The approaches of cause-specific hazards and incorrectly assumed correlations do not have standard errors for tau since tau was not estimated but rather explicitly assumed.

Second, the mean estimates of the regression coefficients $\beta_{T}$ and $\beta_{C}$ were evidently unbiased using the proposed method of correlation estimation, compared to the other three scenarios. The mean estimates of $\widehat{\beta}_{T} = -0.482$ and $\widehat{\beta}_{C} = 0.187$ were close to the true values of $-0.5$ and 0.2, compared to the largely over-estimated values under independence and an incorrectly weaker correlation of 0.3, as the mean percentage errors clearly demonstrate. The beta estimates under an incorrectly stronger correlation of 0.9 between $T$ and $C$ became under-estimated as $\widehat{\beta}_{T} = -0.436$ and $\widehat{\beta}_{C} = -0.130$ with even the direction of association being reversed for $\beta_{C}$. The absolute deviation from the true regression coefficients increased as the assumed correlation further deviated from the truth, underscoring the importance of accurately estimating dependence in dependently censored survival data.

\begin{table}[H]
\centering
\caption{Baseline characteristics of the AIDS Clinical Trials Group (ACTG) 175 dataset ($N=1{,}054$ patients) in total and by treatment arms}
\label{tab:tableS3}
\begin{threeparttable}
\small
\begin{tabular}{lcccc}
\toprule
Variables & Total & Mono & Combo & P-value$^{\text{a}}$ \\
\midrule
Overall N (\%) & 1,054 (100) & 532 (50.5) & 522 (49.5) & -- \\
Age, mean (SD) & 35.2 (8.77) & 35.2 (8.85) & 35.2 (8.70) & 0.994 \\
\midrule
Gender, N (\%) & & & & 0.458 \\
\quad Male & 866 (82.2) & 432 (81.2) & 434 (83.1) & \\
\quad Female & 188 (17.8) & 100 (18.8) & 88 (16.9) & \\
\midrule
Race, N (\%) & & & & 0.329 \\
\quad White & 760 (72.1) & 376 (70.7) & 384 (73.6) & \\
\quad Other & 294 (27.9) & 156 (29.3) & 138 (26.4) & \\
\midrule
History of IV drug use, N (\%) & & & & 0.344 \\
\quad Yes & 136 (12.9) & 63 (11.8) & 73 (14.0) & \\
\quad No & 918 (87.1) & 469 (88.2) & 449 (86.0) & \\
\midrule
Hemophilia, N (\%) & & & & 0.927 \\
\quad Yes & 85 (8.1) & 42 (7.9) & 43 (8.2) & \\
\quad No & 969 (91.9) & 490 (92.1) & 479 (91.8) & \\
\midrule
Baseline CD4 count, mean (SD) & 351 (122) & 349 (130) & 353 (114) & 0.552 \\
\midrule
Prior antiretroviral therapy, N (\%) & & & & 0.761 \\
\quad Yes & 618 (58.6) & 309 (58.1) & 309 (59.2) & \\
\quad No & 436 (41.4) & 223 (41.9) & 213 (40.8) & \\
\midrule
Disease symptoms, N (\%) & & & & 0.530 \\
\quad Yes & 185 (17.6) & 89 (16.7) & 96 (18.4) & \\
\quad No & 869 (82.4) & 443 (83.3) & 426 (81.6) & \\
\midrule
Primary endpoint, N (\%) & & & & $<0.001$ \\
\quad Yes & 284 (26.9) & 181 (34.0) & 103 (19.7) & \\
\quad No & 770 (73.1) & 351 (66.0) & 419 (80.3) & \\
\midrule
Follow-up time (years), mean (SD) & 858.2 (302.9) & 801.2 (326.9) & 916.2 (264.2) & $<0.001$ \\
\bottomrule
\end{tabular}
\begin{tablenotes}[flushleft]\footnotesize
\item Abbreviations: ACTG, AIDS Clinical Trials Group Study; Combo, combination therapy of AZT plus didanosine; IV, intravenous; Mono, AZT monotherapy; SD, standard deviation.
\item[$^{\text{a}}$] P-values were calculated with ANOVA for continuous variables and the Chi-squared test for categorical variables.
\end{tablenotes}
\end{threeparttable}
\end{table}

\bibliographystyle{mybibstyle}
\bibliography{references}

\end{document}